\documentclass[aps,prl,twocolumn,superscriptaddress,groupedaddress]{revtex4}
\usepackage{graphicx}  
\usepackage{dcolumn}   
\usepackage{bm}        
\usepackage{amssymb}   
\usepackage{xcolor}	   

\begin{document}



\title{Origin of the butterfly magnetoresistance in ZrSiS}
\author{J.A.~Voerman} \affiliation{MESA+ Institute for Nanotechnology, University of Twente, 7500 AE Enschede, The Netherlands}
\author{L.~Mulder} \affiliation{MESA+ Institute for Nanotechnology, University of Twente, 7500 AE Enschede, The Netherlands}
\author{J.C. de Boer}\affiliation{MESA+ Institute for Nanotechnology, University of Twente, 7500 AE Enschede, The Netherlands}
\author{Y.~Huang} \affiliation{Van der Waals-Zeeman Institute, University of Amsterdam, Science Park 904, 1098 XH Amsterdam, The Netherlands}
\author{L.M.~Schoop} \affiliation{Department of Chemistry, Princeton University, Princeton, New Jersey 08544, United States}
\author{Chuan Li} \affiliation{MESA+ Institute for Nanotechnology, University of Twente, 7500 AE Enschede, The Netherlands}
\author{A.~Brinkman}\affiliation{MESA+ Institute for Nanotechnology, University of Twente, 7500 AE Enschede, The Netherlands}
\vskip 0.25cm
\date{\today}

\begin{abstract}
ZrSiS has been identified as a topological material made from non-toxic and earth-abundant elements. Together with its extremely large and uniquely angle-dependent magnetoresistance this makes it an interesting material for applications. We study the origin of the so-called butterfly magnetoresistance by performing magnetotransport measurements on four different devices made from exfoliated crystalline flakes. We identify near-perfect electron-hole compensation, tuned by the Zeeman effect, as the source of the butterfly magnetoresistance. Furthermore, the observed Shubnikov-de Haas oscillations are carefully analyzed using the Lifshitz-Kosevich equation to determine their Berry phase and thus their topological properties. Although the link between the butterfly magnetoresistance and the Berry phase remains uncertain, the topological nature of ZrSiS is confirmed.
\end{abstract}

\maketitle

\section{I. Introduction}
ZrSiS belongs to a class of electronic materials known as nodal-line semimetals \cite{neupane2016,schoop2016}. These materials are distinct from topological insulators (TI) and Dirac/Weyl semimetals (DSM/WSM), because the bands cross more than once in momentum space to form lines or circles of Dirac nodes. \cite{burkov2011,bzdusek2016,fang2016,chen2015,yang2018}. A number of studies have reported on the enormous, nonsaturating magnetoresistance (MR) that ZrSiS displays \cite{ali2016,lv2016,wang2016,singha2016,zhang2018}, a feature that makes it a good candidate for sensor applications. The general usability of ZrSiS is supported by its stability and elemental abundance. The material contains three non-toxic and ubiquitous elements and shows no signs of degradation in ambient conditions \cite{sankar2017,schoop2016}. Not only is the MR in ZrSiS extremely large, it has a peculiar angular dependence as well, showing a maximum MR when the angle between the applied magnetic field and the sample surface is $45^{\circ}$, instead of a perpendicular field as one would expect. When plotted, the MR has a distinct butterfly shape and so the effect has been christened \textit{butterfly magnetoresistance} \cite{ali2016,wang2016,hu2016,zhang2018}.
Although numerous studies have addressed the butterfly MR both from an experimental as well as a theoretical angle, where the intricate band structure from which the transport properties arise was examined \cite{neupane2016,schoop2016,yang2018,topp2017,rudenko2018}, a clear link between the two has not yet been established. Due to the high mobility of the charge carriers in ZrSiS, studying quantum oscillations provides us with a great opportunity to understand the unique transport properties of this material, as illustrated by the number of reports on quantum oscillations in ZrSiS \cite{pezzini2018,matusiak2017,hu2017,li2018}. In this article we study the origin of the butterfly MR. To this end we determine the charge carrier density and mobility of different ZrSiS thin flakes and discuss how this relates to MR effects. This is followed by an analysis of the measured Shubnikov-de Haas (SdH) oscillations, which together lead to a possible explanation of the butterfly MR. The SdH oscillations also provide an opportunity to study whether or not the nodal line in ZrSiS indeed exhibits topological properties.

\section{II. Experimental details}
\begin{figure*}
\centering
\includegraphics[width=0.85\textwidth]{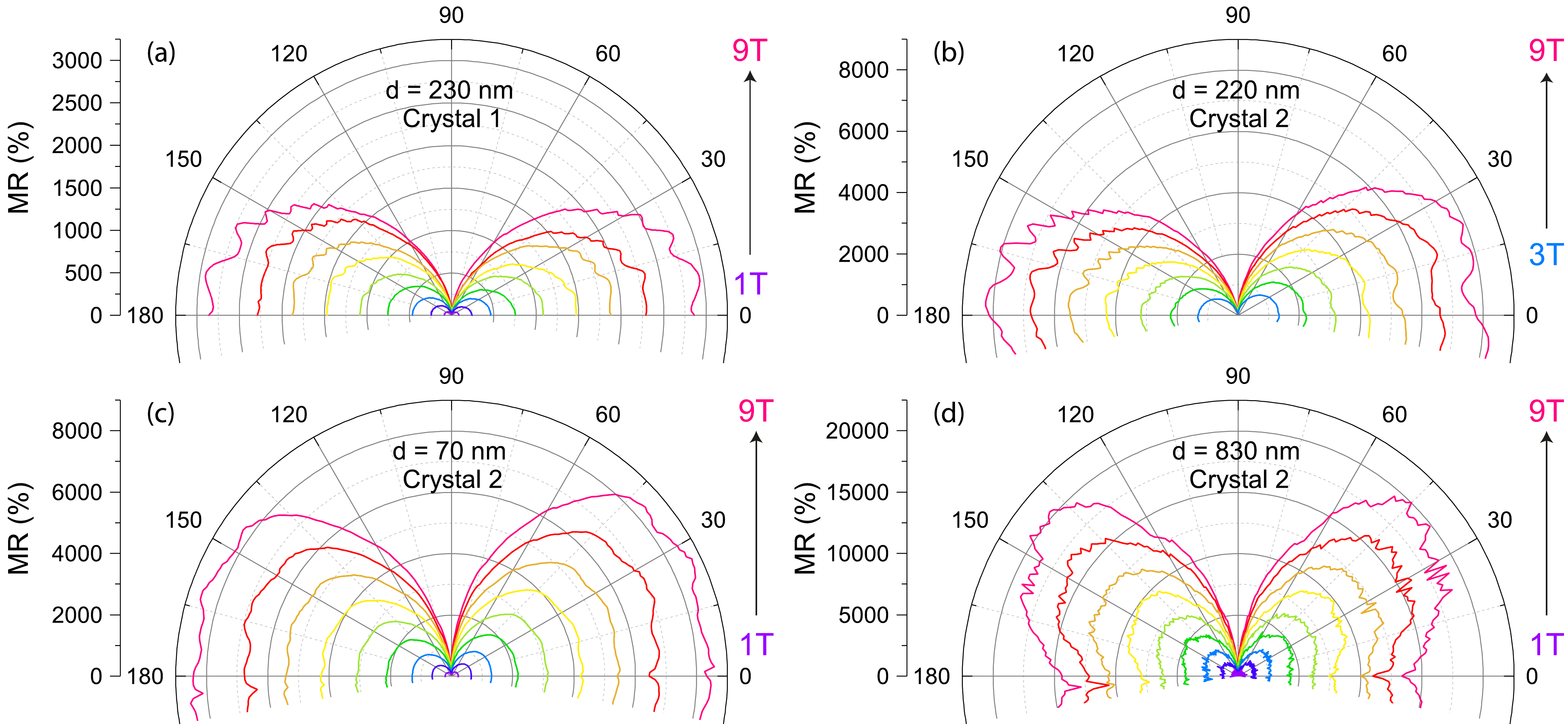}
\caption{\label{fig:allbutterflies}Angle-dependence of the MR of two samples made from different ZrSiS crystals, where $0^{\circ}$ is out-of-plane. Device (a) is made from a 230 nm thick flake of crystal 1. The MR does not exceed 2500\% and the MR has a maximum at $0^{\circ}$. (b) ADMR of a 220 nm thick flake of crystal 2. The MR has a maximum around $20^{\circ}$. (c) ADMR of a 70 nm flake of crystal 2 showing a flat maximum MR of 8000\% between 0$^{\circ}$ and $45^{\circ}$. (d) ADMR of the 830 nm thick flake of crystal 2. The MR reaches values of 20.000\%. The ADMR is clearly butterfly-shaped, having a maximum at $45^{\circ}$.}
\end{figure*}

To create the devices, we have exfoliated flakes of two different ZrSiS crystals. The first ZrSiS single crystals were prepared by a chemical vapor transport method. The stoichiometric mixture of Zr, Si, and S powder was sealed in a quartz ampoule with iodine as transport agent (20 mg cm$^3$). The quartz ampoule was placed in a tube furnace with a temperature gradient from 1000$^\circ$C to 900$^\circ$C for 10 days. Crystal 2 was grown with the same method, but in a carbon coated ampoule to avoid reaction of the samples with the quartz. This crystal was subsequently annealed at 600$^\circ$C for two weeks. The crystal structure of ZrSiS exhibits a natural cleavage plane perpendicular to the c-axis. All exfoliated flakes therefore show the (001)-plane as their top surface \cite{schoop2016}. The flakes were deposited on a silicon substrate coated with 300 nm SiO$_2$. After determining the thickness of the flake using atomic force microscopy, gold contacts were designed and deposited on the flakes using e-beam lithography and RF sputter deposition. In this work we describe measurements performed on four standard six-probe devices made in this fashion. The first of these uses a flake from crystal 1, the other three are made from flakes from crystal 2. We have studied further devices based of crystal 1 including a bulk devices contacted with gold wires and silver epoxy. All measurements on the devices of crystal 1 showed similar results so the results of only one device will be presented here.\\

The devices were cooled down to 2 K in a Quantum Design Physical Property Measurement System (PPMS) on an insert that can be rotated 180 degrees in a maximum magnetic field of 9 T while recording the longitudinal and transversal voltage in the device in a four-probe configuration. The measurements were performed using a 100 $\mu$A current. Here $0^{\circ}$ and $180^{\circ}$ represent an out-of-plane magnetic field. 

\begin{figure*}[t!]
\centering
\includegraphics[width=.7\textwidth]{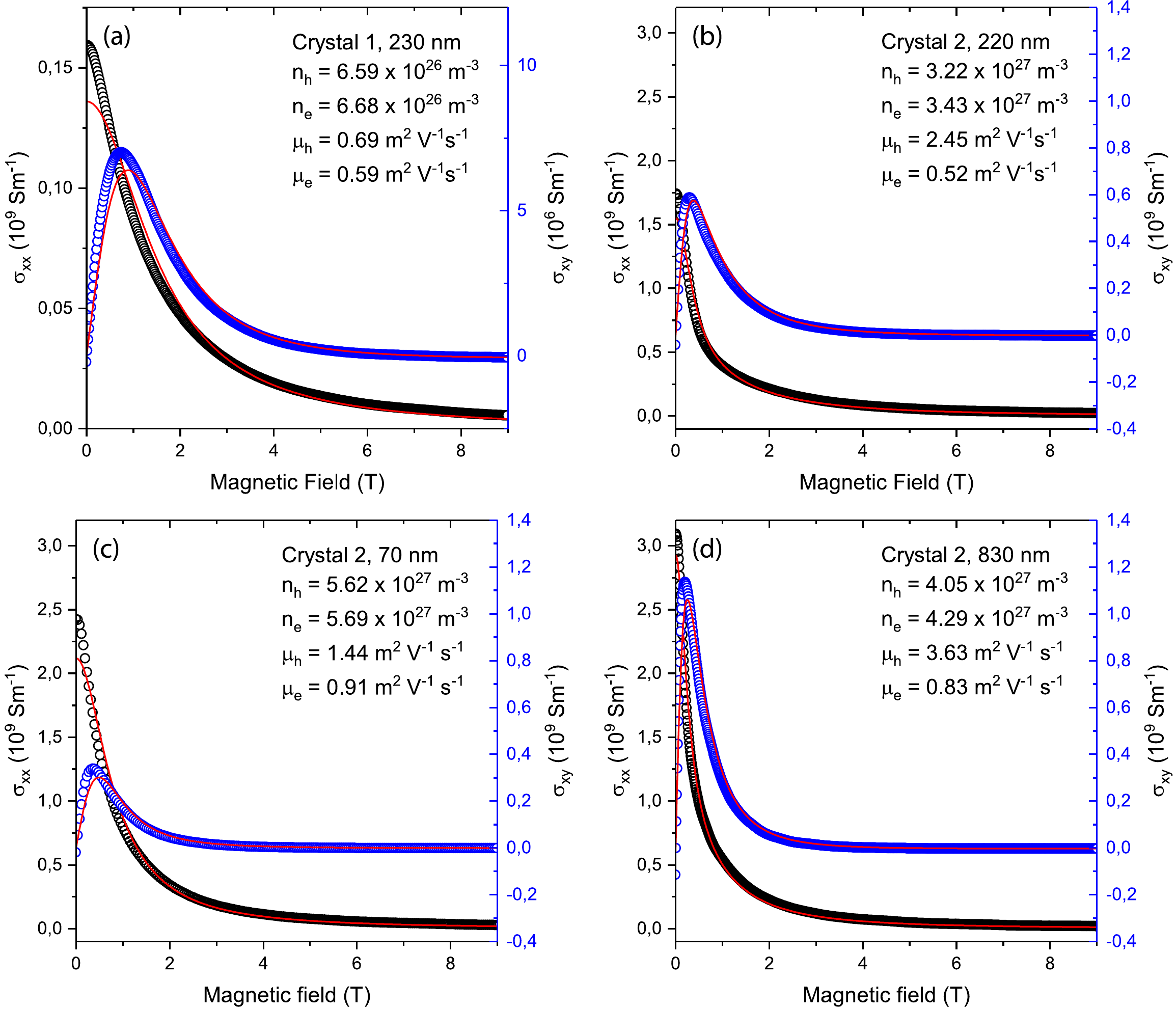}
\caption{\label{fig:2bandfit}Two-band Drude fits (red lines) of the longitudinal (black circles) and transverse (blue circles) conductivity for two devices measured using a bias current of 100 $\mu$A. All fitting parameters are indicated in the top-right corner. (a) Measurement data of the 230 nm thick flake of crystal 1. (b) Measurement data of the 220 nm thick flake of crystal 2. (c) Measurement data of the 70 nm thick flake of crystal 2. (d) Measurement data of the 830 nm thick flake of crystal 2.}
\end{figure*}

\section{III. Angle-dependent magnetoresistance}
Figure \ref{fig:allbutterflies} shows the angle-dependent magnetoresistance (ADMR) of all four devices. The MR is calculated as $100\% \times [R(B)-R(0)]/R(0)$. The results for the different devices differ on two points: the magnitude of the MR and the shape of the ADMR. Although slightly obscured by the SdH oscillations that are present for high magnetic fields and angles close to 0$^{\circ}$, the maximum MR in figure \ref{fig:allbutterflies}(a) is observed at $0^{\circ}$ and $180^{\circ}$, as one would expect for a regular homogeneous conductor. The maximum MR in figure \ref{fig:allbutterflies}(d) lies around $45^{\circ}$, a feature that is known as the butterfly MR. The size of the MR varies by a factor of 8 between the two graphs. The device made from the first crystal of ZrSiS shows an MR of about 2500\%, which can be seen in figure \ref{fig:allbutterflies}(a), whereas a device from the second crystal of ZrSiS shows an MR exceeding 20.000\%. The ADMR of the 220 nm and 70 nm thick devices of crystal 2 is plotted in figure \ref{fig:allbutterflies}(b) and \ref{fig:allbutterflies}(c) respectively. The device based on the 830 nm thick flake, shown in \ref{fig:allbutterflies}(d), exhibits the clearest butterfly MR. The device based on the 70 nm thick flake also shows signs of butterfly MR, but we find that the 220 nm thick flake device does not exhibit any butterfly MR, indicating that the observability of the butterfly MR is not thickness dependent per se. The maximum MR of both the 70 nm and 220 nm thick devices is around 8000\%, 2.5 times lower than for the 830 nm thick device. Futhermore, this maximum is not centered around at $45^{\circ}$, as it is in figure \ref{fig:allbutterflies}(d). For the 220 nm thick device (figure \ref{fig:allbutterflies}(b)) the maximum MR at $45^{\circ}$ is absent altogether and the 70 nm thick device  (figure \ref{fig:allbutterflies}(c)) does reach a maximum at $45^{\circ}$, but keeps this value towards $0^{\circ}$ (perpendicular field) instead of decreasing. It effectively displays a combination of the regular ADMR and the butterfly MR. Comparing all measurements we can conclude that crystal 2 exhibits a butterfly MR similar to what was observed by Ali \textit{et al.} \cite{ali2016}. 


Having observed a butterfly MR in some, but not in all samples, provides a great opportunity to study this anomalous version of the extremely large MR. Such a large increase in resistance can be found in other materials as well and can be caused by a number of effects. One effect is Abrikosov's linear MR, which is present in situations where only the lowest Landau level is occupied \cite{Abrikosov1969,Abrikosov2000}. But while the charge carrier mobilities of the ZrSiS may be high enough for this effect to arise, the carrier density in ZrSiS is rather high to be in the extreme quantum limit and the MR has a strong parabolic component. It has been shown that inhomogeneous thin films can be modeled as a resistor network and that these can exhibit very large MR of mixed linear and parabolic character \cite{Meera2003,Meera2005}. However, with the mobilities typically found for ZrSiS this situation seems unlikely. Another candidate source for the large MR in ZrSiS is a near-perfect balance between electron and hole densities \cite{lv2018,wang2016,matin2018}. Even though the assumptions made within the Drude model are not fully justifiable for an anisotropic system like ZrSiS, we use the Drude two-band model to obtain approximate charge carrier densities and mobilities for the system to provide further insight. In this model the longitudinal resistivity is given by

\begin{eqnarray}
\rho_{xx}(B) = \frac{1}{e}\frac{(n_h\mu_h+n_e\mu_e) + (n_h\mu_e+n_e\mu_h)\mu_h\mu_eB^2}{(n_h\mu_h+n_e\mu_e)^2 + (n_h-n_e)^2\mu_h^2\mu_e^2B^2}.
\end{eqnarray}

\noindent For large fields the rightmost terms of the numerator and denominator dominate the leftmost terms, so we can focus on former. From the denominator it is clear that when the hole and electron density become equal the resistivity has a maximum. The exact value of the resistance still depends on the mobilities of the two bands. Figure \ref{fig:2bandfit} shows the fit of the two-band Drude model to the longitudinal and transverse conductivities of the devices for which the ADMR was presented in figure \ref{fig:allbutterflies}. Figure \ref{fig:2bandfit}(a) shows the data and the best fit to the data of the device made from crystal 1, which exhibits the lowest MR and no butterfly MR. Figure \ref{fig:2bandfit}(b-d) show the data and fits on the 70 nm, 220 nm and 830 nm thick devices made from crystal 2. In general, samples from crystal 2 show more anisotropic ADMR and larger MR than the samples taken from crystal 1. The Drude fits in figure \ref{fig:2bandfit} tell us that in all samples, from both crystals, the electron and hole density are nearly compensated, indicated by the $\sigma_{xy}$ that goes to zero for large magnetic fields. These results imply that electron-hole compensation is the mechanism that generates the enormous MR in ZrSiS. 
As has previously been shown in other Dirac materials such as MoSi$_2$ \cite{matin2018} and WTe$_2$ \cite{wang20162,rhodes2015}, the Zeeman effect can influence electron-hole compensation. It may be that through the same mechanism, the Zeeman effect is also responsible for maximum electron-hole compensation in the anisotropic electronic structure of ZrSiS at a $45^{\circ}$ magnetic field angle.\\


In general, rotating the device normal away from the direction of a magnetic field does not alter the carrier density. However, besides the formation of cyclotron orbits, typically associated with MR, the magnetic field induces a Zeeman shift in the electronic structure. In a crowded and complex band structure such as that of ZrSiS, the Zeeman effect can slightly alter the size and shape of the Fermi pocket and thereby alter the net carrier density of the individual pockets. As the carrier density - and the effects of small changes in it - highly depends on the exact crystal parameters, this would also explain why the butterfly-shaped MR is not observed in all devices. \\

Besides the carrier density, magnetic fields can also tune the carrier mobility and thereby contribute to the large MR in semimetals \cite{fauque2018}, providing another possible way to cause the observed butterfly MR. Unfortunately, for other magnetic field directions than parallel to the sample normal, the Drude model approximations become increasingly invalid so that we cannot test these hypotheses through Drude two-band fits. The poor fitting results of the Drude two-band model for magnetic field angles other than parallel to the $c$-axis, are likely the result of off-diagonal matrix elements in the conductivity tensor. In a recent work, Novak \textit{et al.} \cite{novak2019} numerically calculate the conductivity for Fermi surfaces obtained through first-principle density functional theory and find that indeed only along high symmetry crystal axes, the electron and hole contributions can be considered as parallel conduction channels.\\

\section{IV. Quantum Oscillations}
\begin{figure}
\centering
\includegraphics[width=0.48\textwidth]{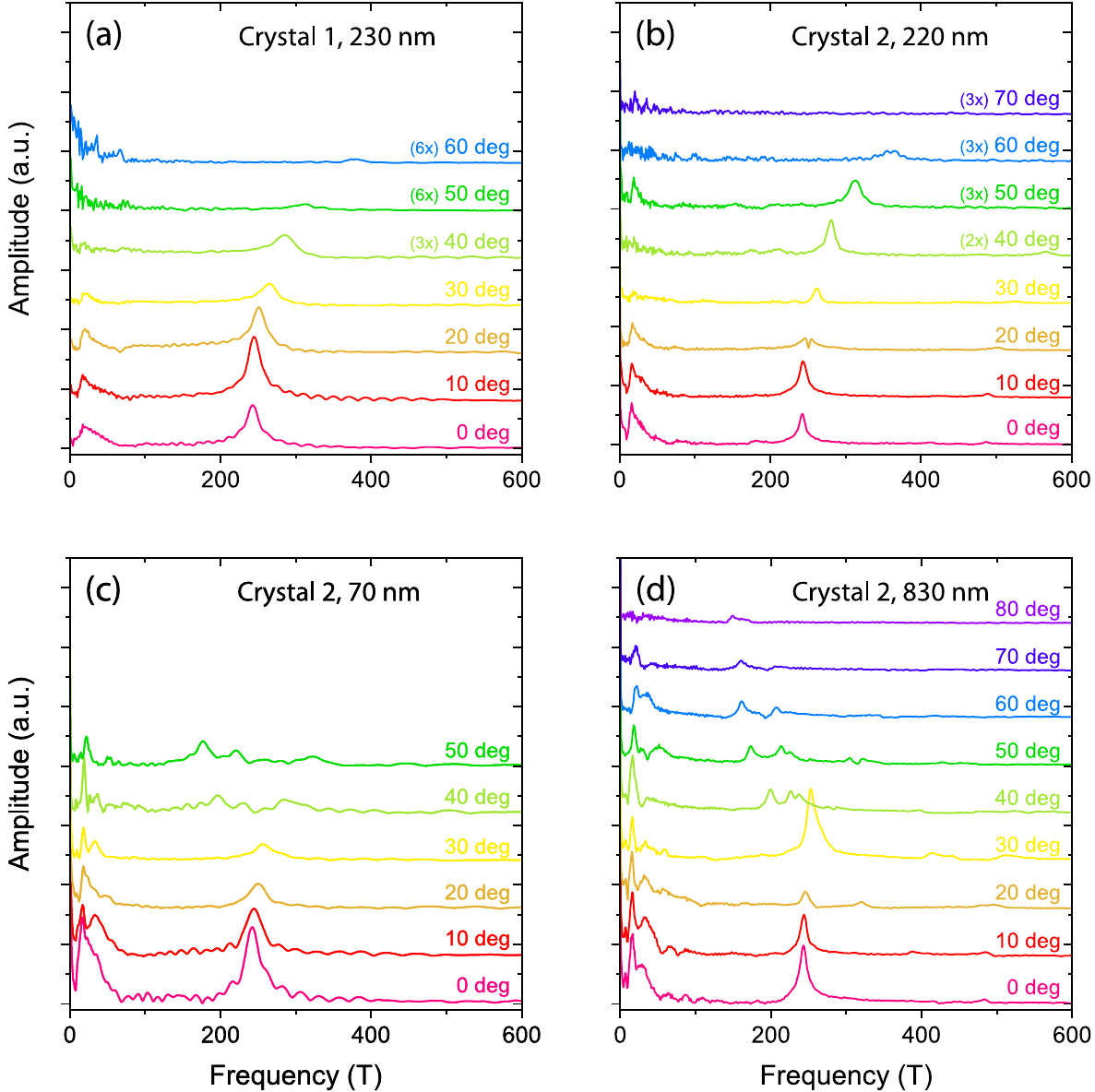}
\caption{\label{fig:sdh}(a)-(d) Fourier spectra of the SdH oscillations measured at different sample orientations. The data is offset for clarity. The angle between the sample and the magnetic field is indicated in the graphs for each line. The thickness of the device on which data has been taken is indicated on the top of each graph. Figures (a) and (b) show the 243 T peak moving towards higher frequencies with increasing angle. These devices do not exhibit the butterfly MR. Figures (c) and (d) each show two peaks moving towards lower frequencies for angles larger than $30^{\circ}$. These devices do show butterfly MR.}
\end{figure}

To extract the SdH oscillations for the longitudinal MR, the measured voltage was symmetrized. The background was subtracted from the signal by smoothing the oscillations. Figure \ref{fig:sdh} shows the fast Fourier transforms of all the oscillations. The data has been offset for clarity and for some larger angles the data has been amplified for visibility, as shown by a factor in front of the angle indication.
Two distinct peaks can be identified at the perpendicular orientation ($0^{\circ}$), located at frequencies of 16 T and 243 T. These SdH frequencies are well known in literature \cite{ali2016,matusiak2017,sankar2017,zhang2018}, and their position as a function of the angle between the sample and the magnetic field correlates with the presence of the butterfly MR. This link is clearly visible by comparing figures \ref{fig:allbutterflies} and \ref{fig:sdh}, and has been observed in literature \cite{ali2016}. The oscillation with a frequency of roughly 16 T does not move as a function of the angle, indicating that the Fermi pocket it corresponds to is spherical. The Onsager relation tells us that the extremal cross-section, $S$, of the Fermi pocket that is responsible for an oscillation \cite{onsager1952}, is given by

\begin{eqnarray}
S = \frac{2\pi e F}{\hbar},
\end{eqnarray}

\noindent where $F$ is the frequency of the oscillation in Tesla. For a 16 T oscillation this gives a reciprocal area of $1.53 \times 10^{-3} $\AA$^{-2}$. The peak that starts at $F = 243$ T in all panels of figure \ref{fig:sdh} corresponds to a reciprocal area of $2.32 \times 10^{-2} $\AA$^{-2}$. Both these frequencies are in good correspondence with literature \cite{ali2016,pezzini2018,matusiak2017}. The 3D Brillouin zone (BZ) of ZrSiS is a fourfold symmetric structure that has two distinct features. In the $\Gamma$-Z direction one finds elongated tubular shapes that extend in the $k_z$ direction. These tubular shapes are hole-pockets and have an extremal cross-section that corresponds well with a SdH frequency of $243$ T. The other main feature is an electron-pocket situated in the $\Gamma$-M direction. These have a larger extremal cross-section than the tubes, corresponding to a SdH frequency between $550$ T and $600$ T. The SdH oscillations from this pocket have been studied in literature \cite{pezzini2018,matusiak2017}, but the oscillation period is too large to reliably measure in our set-up and the electron mobility is in general lower than the hole mobility, making the SdH oscillations from the electrons more difficult to distinguish. Albeit faintly, it can be seen in some of the Fourier spectra, such as the 10$^{\circ}$ lines of figure \ref{fig:sdh}(b) and (d). The smallest reciprocal area, corresponding to the F $=$ 16 T peak, is more difficult to link to the Fermi surface, but has been observed before \cite{ali2016,matusiak2017}. 

The $F(\theta)$ dependence of the $243$ T peak displayed in figures \ref{fig:sdh}(c) and (d) is quite the opposite of that in figures \ref{fig:sdh}(a) and (b). Although figure \ref{fig:sdh}(c) still faintly exhibits a rightmoving peak, both (c) and (d) clearly show that from 30$^{\circ}$ onwards the peak splits and moves to lower frequencies. This is in perfect correlation with the appearance of the butterfly MR as is evidenced by the thinnest device, presented in figure \ref{fig:allbutterflies}(c) and figure \ref{fig:sdh}(c), which shows a mixture of regular and butterfly angle-dependence in both its MR and SdH oscillations. From this comparison it is clear that the hole-pocket plays an important role in the appearance of the butterfly MR. 

Quantum oscillations in materials, such as the SdH oscillations in ZrSiS observed here, can be described with the model created by Lifshitz and Kosevich in 1956 \cite{lifshitz1956,hu2017b}. The Lifshitz-Kosevich (LK) model can be used to determine whether or not that particular path through the BZ is accompanied by a Berry phase, $\phi_B$, which is a clear indication of topological transport. Futhermore, the phase of the quantum oscillations is influenced by the carrier type, the dimensionality of the band, and the dispersion of the band (linear of parabolic). In a recent publication, Li \textit{et al.} have collected all possible phase shifts in the LK model for a topological nodal-line semimetal and warn that claiming non-trivial transport is not so trivial \cite{li2018}. The circular orbit around the hole-populated tube, from which the $243$ T oscillation originates, is described as a quasi-2D tube and therefore associated with a geometrical phase shift of $\delta = 0$. Furthermore, because this orbit encloses a Dirac point it should yield a discernable Berry phase $\phi_B = \pi$ in the quantum oscillations. Hyun \textit{et al.} have studied the Berry phase in materials with a square-net substructure, belonging to the P4/nmm or Pnma space group, and in ZrSiS specifically \cite{hyun2018}. In their work they highlight the tubular hole-pocket as a ``Berry hot spot'', a Fermi surface that encloses a large Berry curvature. Importantly, they note that the Berry phase strongly depends on the Fermi energy and electron-hole asymmetry. Other investigations into the topological nature of this Fermi pocket have yielded a variety of conclusions, including fully trivial \cite{zhang2018}, fully topological \cite{pezzini2018}, Berry phases that are neither $\pi$ nor $0$ \cite{hu2017}, and a Berry phase that changes from trivial to topological for some angles of the magnetic field \cite{ali2016}. These articles have been written before the publication of the theoretical work by Li \textit{et al.} and have all assumed 3D carriers, whereas Li \textit{et al.} note that the 243 T frequency is mostly in the (quasi-)2D regime \cite{li2018}.\\

\begin{figure*}
\centering
\includegraphics[width=\textwidth]{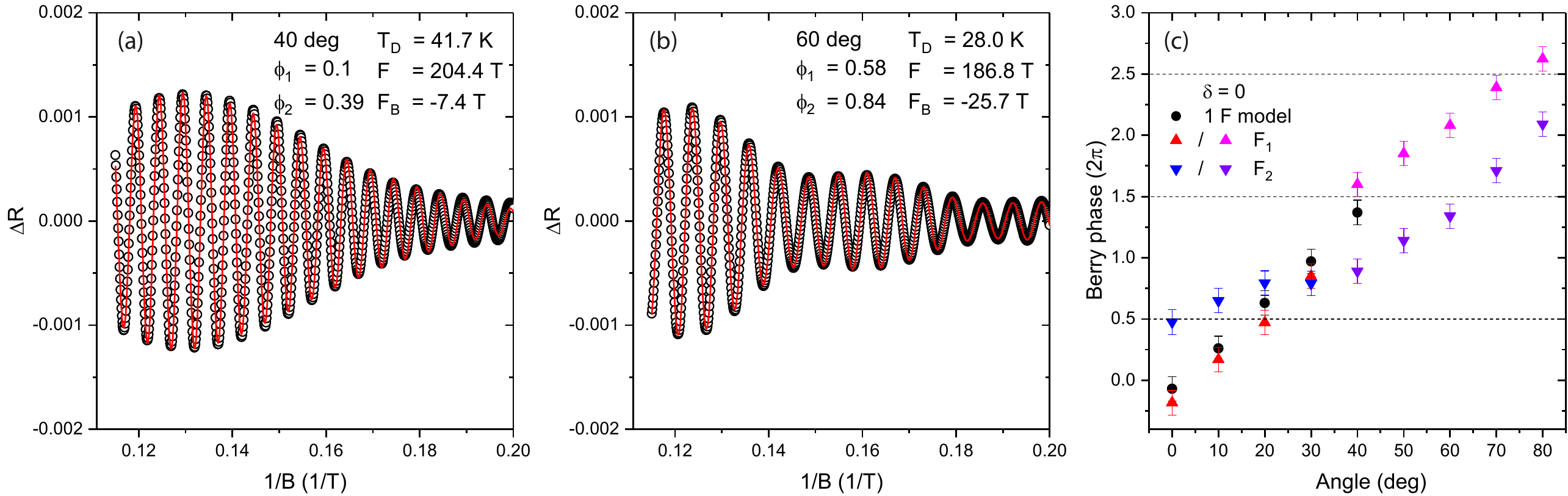}
\caption{\label{fig:berryphases}(a) and (b) show two examples of the LK model applied to the extracted SdH oscillations (black circles) from the measurements on the 830 nm thick device. The fits (red lines) have been made by adding two frequencies to create the beating pattern. All relevant fitting parameters as well as the corresponding angle are indicated in each panel. $F_B$ is the difference between the two used frequencies. (c) Berry phases obtained from the LK fits, using $\delta = 0$. The black circles are derived from fits using only one frequency. Red/magenta triangles show $\phi_B$ of the first frequency. The blue/violet triangles show $\phi_B$ of the second frequency. }
\end{figure*}

Figures \ref{fig:berryphases}(a) and (b) show, as an example, two typical SdH oscillations measured on the device that exhibits clear butterfly MR. Nearly all of the extracted oscillations are characterized by a beating pattern, which is especially prevalent from 40$^{\circ}$ onwards. Therefore, modeling these oscillations using the LK model with one frequency results in poor fitting results. By applying a bandpass filter around the desired frequency some of the beating can be suppressed. Even then, it is not always possible to fit the full extent of the oscillation. Beating is caused by the presence of two similar, but different, frequencies and since there is an obvious beating pattern in these SdH oscillations, a LK model using two frequencies is more appropriate. If the Zeeman effect plays an important role in the formation of the butterfly MR, as was suggested earlier, two nearly identical SdH frequencies are to be expected. The red lines in figures \ref{fig:berryphases}(a) and (b) are the best fits to the data using this two-frequency model and confirm the presence of two distinct frequencies. Zeeman splitting itself does not change the shape of the two (now non-degenerate) pockets, but in the complex BZ of ZrSiS, the Zeeman shift may even deform pockets slightly, which would show up as a beating pattern between the altered and unaltered pocket.\\

All relevant fitting parameters of the two-frequency model are indicated in the top-right of the respective panel. The Berry phases resulting from this analysis have been included in figure \ref{fig:berryphases}(c). The Berry phases obtained by the single-frequency procedure are shown as black circles. The red and magenta triangles belong to one of the frequencies of the two-frequency model, while the blue and violet triangles belong to the other. The phases have been drawn as red and magenta to remind the reader of the fact that for angles of 40$^{\circ}$ and larger there are several peaks in the SdH spectrum (figure \ref{fig:sdh}(d)). Here, we examine the left-most one of the three. The dashed lines serve as a reading guide and indicate $\phi_B = \pi$, which is typically associated with topological states. The main feature of figure \ref{fig:berryphases}(c) is that the obtained $\phi_B$ increases with angle. Although it starts out at the trivial state, for the single-frequency model and first frequency of the two-frequency model, the phases increase, alternating between trivial and topological. For clarity, the Berry phases have been drawn to go up to 5$\pi$, even though in practice $\phi_B$ is 2$\pi$-periodic. At 40$^{\circ}$, the angle where the peaks in the Fourier spectrum split and where the butterfly MR has a maximum in resistance, the Berry phase suddenly makes a jump of 1.5$\pi$ $\pm$ 0.4$\pi$. Generally speaking, it is possible that the Berry phase deviates from the standard cases of zero (trivial) and $\pi$ (topological). These values are only found when the path through the BZ fully encompasses the Dirac node, as for example a spherical pocket does. Any extremal cross-section must then necessarily fully encompass the Dirac node if it is present, hence we find either $\phi_B = 0$ of $\phi_B = \pi$. But one that does not fully enclose the Dirac node thus encircles an arbitrary amount of Berry connection and in such a situation there can be any $\phi_B$ between 0 and $\pi$. This can also be understood in terms of the spin-texture as only the projection of the spin onto the orbital plane makes a full rotation. If there is an additional out-of-plane component, the obtained Berry phase will be somewhere between 0 and $\pi$.\\

From Hyun \textit{et al.} and the comparison of our work to other works in literature we know that the topological properties of ZrSiS are highly sensitive to properties such as the Fermi level and electron/hole ratio \cite{ali2016,zhang2018,pezzini2018,matusiak2017,hyun2018}. The latter could be the link between the Berry phase and the butterfly MR, as the importance of the electron/hole ratio was also identified earlier. Upon tilting the magnetic field one could potentially make two orbits touch and fuse together, instantly increasing the size of the loop, $S$, which would show as a sudden jump in Berry phase such as was observed by Ali \textit{et al.} \cite{ali2016}. Despite the speculative nature of this argument, it could explain the mechanism responsible for the jump in Berry phase at 40$^{\circ}$ in figure \ref{fig:berryphases}(c).

\section{V. Conclusions}
The MR in ZrSiS can safely be labeled extremely large. This large MR can be understood by considering a two-band Drude model where the two bands are equally populated by opposite charge carriers. This near-perfect electron-hole compensation may also be the driving mechanism behind the peculiar butterfly MR, where the MR has a maximum, not for a perpendicular magnetic field, but when the angle between the current and magnetic field is $45^{\circ}$. We argue that the hole and electron densities may become comparable around this angle through the Zeeman effect, and that this creates a maximum in the MR. This mechanism is supported by the observation of beating patterns in the SdH oscillations, which is caused by two pockets with near-identical extremal cross-sections.\\

From careful analysis of the SdH oscillations we conclude that the tubular hole-like Fermi pocket in the $\Gamma-Z$ direction of the BZ is a contributing factor to the butterfly MR, in combination with the larger, neighboring electron pocket. Rotating the samples in a magnetic field allows one to study the size of the extremal cross-section of the pocket as viewed from different angles. Any device that exhibits butterfly MR will show a transition from the extremal cross-section becoming larger with increasing angle to splitting into several distinct areas that become smaller with increasing angle.
It remains uncertain, however, whether the topological properties of this hole-like Fermi pocket play a role in the creation of the butterfly MR or that the near-perfectly electron-hole compensated Drude model is solely responsible for this peculiar effect. By fitting a two-frequency LK model to the extracted SdH oscillations, and taking into account the appropriate phase shift for the Fermi pocket under study, we conclude that the pocket has topological properties as it clearly shows a non-zero Berry phase that increases with angle.\\

\section{Acknowledgements}

This work was financially supported by the European Research Council (ERC) through a Consolidator Grant. L.M. is supported by the Netherlands Organisation for Scientific Research (NWO) through a Vici grant. LMS was supported by NSF through the Princeton Center for Complex Materials, a Materials Research Science and Engineering Center DMR-1420541, and by a MURI grant on Topological Insulators from the Army Research Office, grant number ARO W911NF-12-1-0461.

\end{document}